\documentclass{emulateapj}
\usepackage{amssymb,graphicx,longtable,epsfig,lscape}
\usepackage{wasysym}
\def\msol{\hbox{$\rm\thinspace M_{\odot}\thinspace$}} 
 
\def\etal{{\it et al.\thinspace}}

\newcommand{\be}{\begin{equation}}
\newcommand{\ba}{\begin{eqnarray}}
\newcommand{\ee}{\end{equation}}
\newcommand{\ea}{\end{eqnarray}}

\begin{document}

\title{Prompt Ia Supernovae Are Significantly Delayed} 
\author{Cody Raskin\altaffilmark{1}, Evan Scannapieco\altaffilmark{1}, James Rhoads\altaffilmark{1}, Massimo Della Valle\altaffilmark{2,3,4}} 
\altaffiltext{1}{School of Earth and Space Exploration,  Arizona State University, P.O.  Box 871404, Tempe, AZ, 85287-1404.}  
\altaffiltext{2}{INAF- Osservatorio Astronomico di Capodimonte, Salita Moiariello, 16 -80131, Napoli, Italy}
\altaffiltext{3}{European Southern Observatory--Karl Schwarschild Strasse 2- D-85748 Garching bei M\" unchen, Germany}
\altaffiltext{4}{International Centre for Relativistic Astrophysics Network-Piazzale della Republica 2, Pescara, Abruzzo, Italy}

\begin{abstract}

The time delay between the formation of a population of stars and the onset of type Ia supernovae (SNe Ia) sets important limits on the masses and nature of SN Ia progenitors.   Here we use a new observational technique to measure this time delay by comparing the spatial distributions of SNe Ia to their local environments. Previous work attempted such analyses encompassing the entire host of each SN Ia, yeilding inconclusive results. Our approach confines the analysis only to the relevant portions of the hosts, allowing us to show that even so-called ``prompt'' SNe Ia that trace star-formation on cosmic timescales exhibit a significant delay time of 200-500 million years. This implies that either the majority of Ia companion stars have main-sequence masses less than 3\msol, or that most SNe Ia arise from  double-white dwarf binaries.  Our results are also consistent with a SNe Ia rate that traces the white dwarf formation rate, scaled by a fixed efficiency factor.

\end{abstract}

\keywords{stars: evolution -- supernovae: general}

\section{Introduction}

Type Ia supernovae serve as the primary cosmological standard candles (Colgate 1979; Branch \& Tammann 1992; Phillips 1993), due to their extremely regular light curves (Pskovskii 1977; Phillips 1993).  However, this regularity need not imply that they share a common progenitor.  Indeed, one striking bimodality is that the brightest SNe Ia occur in star-forming galaxies, and the dimmest  SNe Ia occur in galaxies with little star-formation (Hamuy \etal 1996; Howell 2001; van den Bergh \& Filippenko 2005). This points to an evolution of SN Ia progenitors, and in fact, while several parameterizations of the SNe Ia rate exist (Greggio \& Renzini 1983; Kobayashi \etal 2000; Greggio 2005), current data is well fit by a two-component parameterization (Mannucci \etal 2005; Scannapieco \& Bildsten 2005), which takes the form  
\be
	SNR_{\mathrm{Ia}}(t) = A{M_*(t)} +B{\dot{M}_*(t)}.
\ee
Here, we refer to the  A-component, which  is proportional to the total stellar mass of the host, as \textit{delayed}, and to the  B-component, which is  is proportional to the instantaneous star-formation rate, as \textit{prompt}.  Note that these words have been occasionally used slightly differently in the literature (Greggio \& Renzini 1983; Mannucci \etal 2006), assigning SNe Ia as prompt or delayed/tardy based on timescales, rather than the 
parameterization in eq.\ (1).

Despite the usefulness of eq.\ (1) on long time scales, there are several reasons to believe that the prompt component should not be pinned to the instantaneous star-fomation rate.  If core-collapse supernovae (SNcc) and prompt SNe Ia occured simultaneously, the oldest stars in the Milky Way would  be much more strongly enriched with iron group elements than observed (McWilliams 1992; Scannapieco \& Bildsten 2005). Furthermore, unlike SNcc, SNe Ia can not arise before stars evolve to form white-dwarfs, which takes at least 40 Myrs. Thus, it is likely that the prompt component exhibits its own delay time, $\tau$, such that
\be
	SNR_{\mathrm{Ia}}(t) = A{M_*(t)} +B{\dot{M}_*(t-\tau)},
\label{eq:ratedelay}
\ee
where current constraints, which use the global properties of host galaxies, place an indirect upper limit of $\tau \leq 500$ Myr on this timescale (Dahlen \etal 2004; Gal-Yam \& Maoz 2004; Mannucci \etal 2005; Scannapieco \& Bildsten 2005; Sullivan \etal 2006; Dilday \etal 2008). Galaxies older than this age have fewer SNe Ia, suggesting the characteristic delay time of the prompt component cannot be longer than 500 Myr. An alternative model is that SNe Ia occur at a rate that is directly proportional to the white dwarf formation rate (WDFR). In particular, Pritchet \etal (2008) argue for a uniform conversion of newly formed white dwarfs into type SNe Ia at an efficiency of $\approx$1\%. This would fix the Ia rate as being proportional to the stellar death rate of stars less than $8\msol$. 

In this paper, we employ a new observational technique for measuring the spatial distributions of SNe Ia and compare the results to analytical models with the goal of constraining the delay time, $\tau$, for the B-component in the A+B formalism, and the feasability of the white dwarf formation rate (WDFR) model discussed below. The structure of this work is as follows. In \S 2, we review previous work constraining the progenitors of other classes of transients, and build on these techniques to develop a new analysis applicable to type Ia SNe. In \S 3, we review the construction of the analytical host model for comparison from Raskin \etal (2008), including modifications necessary to make accurate comparisons to the results of our new observational technique.  We present our results of the analysis and comparisons in \S 4, and conclusions are given in \S 5.

\section{Data and Analysis}

\subsection{Previous work}

Fruchter \etal (2006) developed an observational approach in which, for each host galaxy, they computed the fraction of the total light (or photon counts) in all pixels fainter than the pixel containing a transient. By compiling a sample of such measurements into a cumulative histogram, they demonstrated that long-duration gamma-ray bursts are more likely than SNcc to cluster in the brighter regions of a galaxy. Kelly \etal (2008) expanded upon this analysis to distinguish between SNcc subtypes Ic, Ib, and type II (see also Anderson \& James 2008). 

However, this study showed no difference between SNcc and SNe Ia, even though SNcc arise from massive, short-lived stars rather than from white dwarfs that form from longer-lived stars.  In fact, both populations followed the same distribution as that of the $g$-band light. In the core-collapse case, this is because young stars lead both to supernovae and the brightest regions. On the other hand, SNe Ia are not likely to arise from ongoing star formation, but they nevertheless share the same radial exponential profile as the $g$-band light, which is caused by a radially-decreasing density of stars. This radial gradient is large enough to obscure any signal caused by a delay in the prompt component (Raskin \etal 2008).

\subsection{Doughnut Method}

What is needed is a procedure that correlates  SNe Ia with the properties of nearby regions, rather than with the host as a whole. In a spiral host, the ideal method for constraining SN Ia progenitors would be to measure the relative brightness of pixels within annuli.   In this case, as the density wave of star-formation moves around each annulus, SNe Ia would appear behind it at a characteristic surface brightness determined by the level to which a stellar population fades away in the $g$-band before SNe Ia appear.  The $g$-band is ideal for this analysis as  it fades away on the order of 100 Myrs, a similar time as that of white dwarf formation.  However, observations are never perfect, and observing a single annulus of a spiral host is subject to complications such as spurs, knots, and gaps, as well as the fact that stars rarely follow circular orbits.

\begin{figure}[!ht]
\centering
\includegraphics[width=7cm,height=7cm]{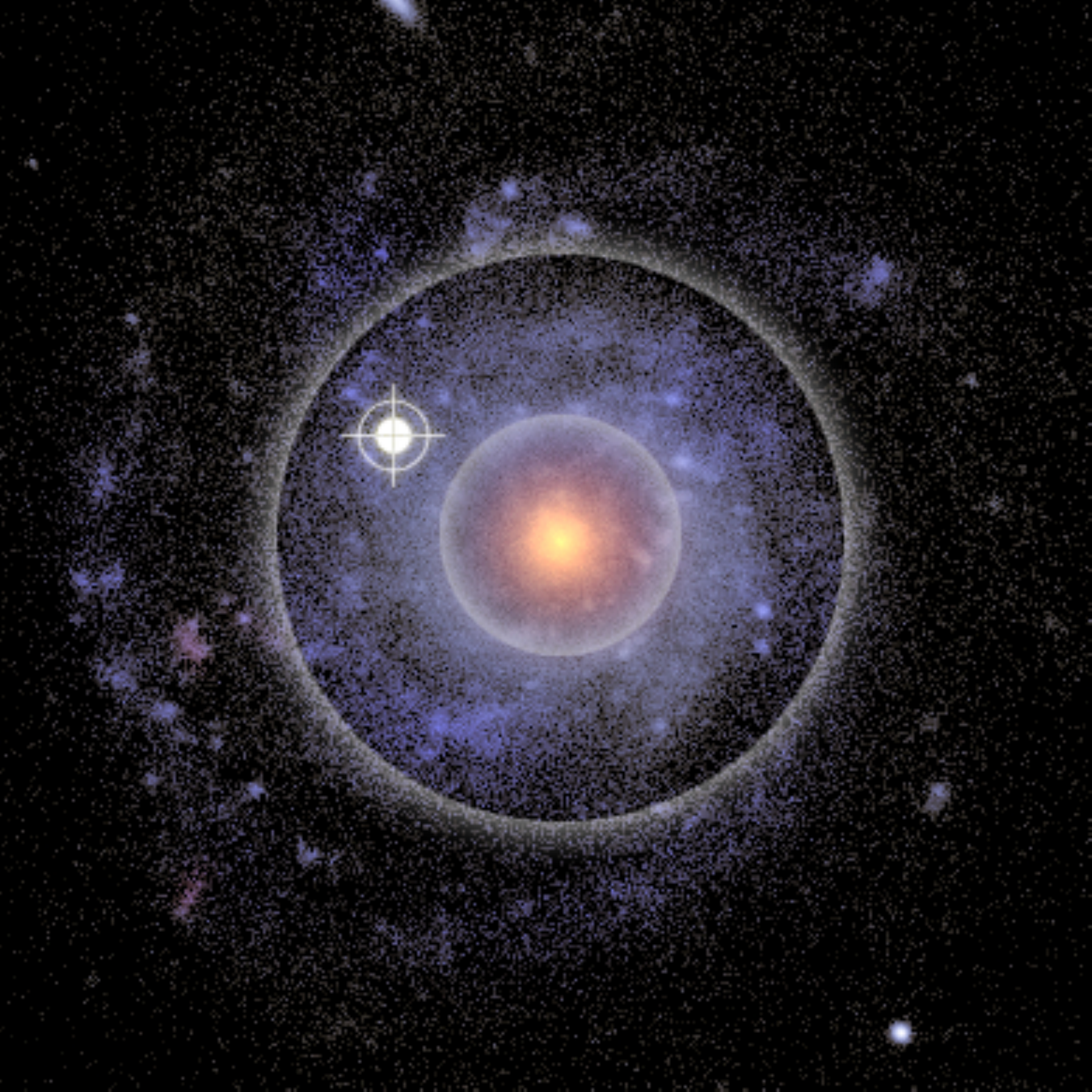}
\caption{A sample image showing a spiral host, NGC 3892, with a doughnut constraint,
	having a width of one-half the scale radius of the galaxy.
	The highlighted region has a small radial gradient, but still samples an
	area in which stars have an appreciable spread in ages reflected in their $g$-band 
		surface brightnesses. The cross marks the location of type Ia supernova 1963J.\\}
\label{fig:Doughnut}
\end{figure}

We call our solution to this problem the \textit{doughnut method}, and it builds directly on the method described in Fruchter \etal (2006). The idea is to expand an annulus radially by some small but appreciable radius, so as to encompass enough of the host's morphological peculiarities to have a good representative sample,
yet narrow enough to represent local variations in the host light. Figure 1 illustrates this concept.

\subsection{Sample Selection}
Our sample was drawn from the  the Padova-Asiago Supernova Catalogue (Barbon \etal 1999),  by selecting those events occurring within $z<0.07$ spiral hosts with Sloan Digital Sky Survey (SDSS) $g$-band images. While this yielded 98 SNe Ia, we also removed  those that occurred within $2''$ of a foreground star or within a galaxy inclined more than  $60^\circ$ or involved in a merger.  This resulted in 50 usable images, and as a control sample,  we also selected all SNcc meeting these criteria, which resulted in 74 usable images. 

\begin{figure}[!ht]
\centering
\includegraphics[width=7cm,height=7cm]{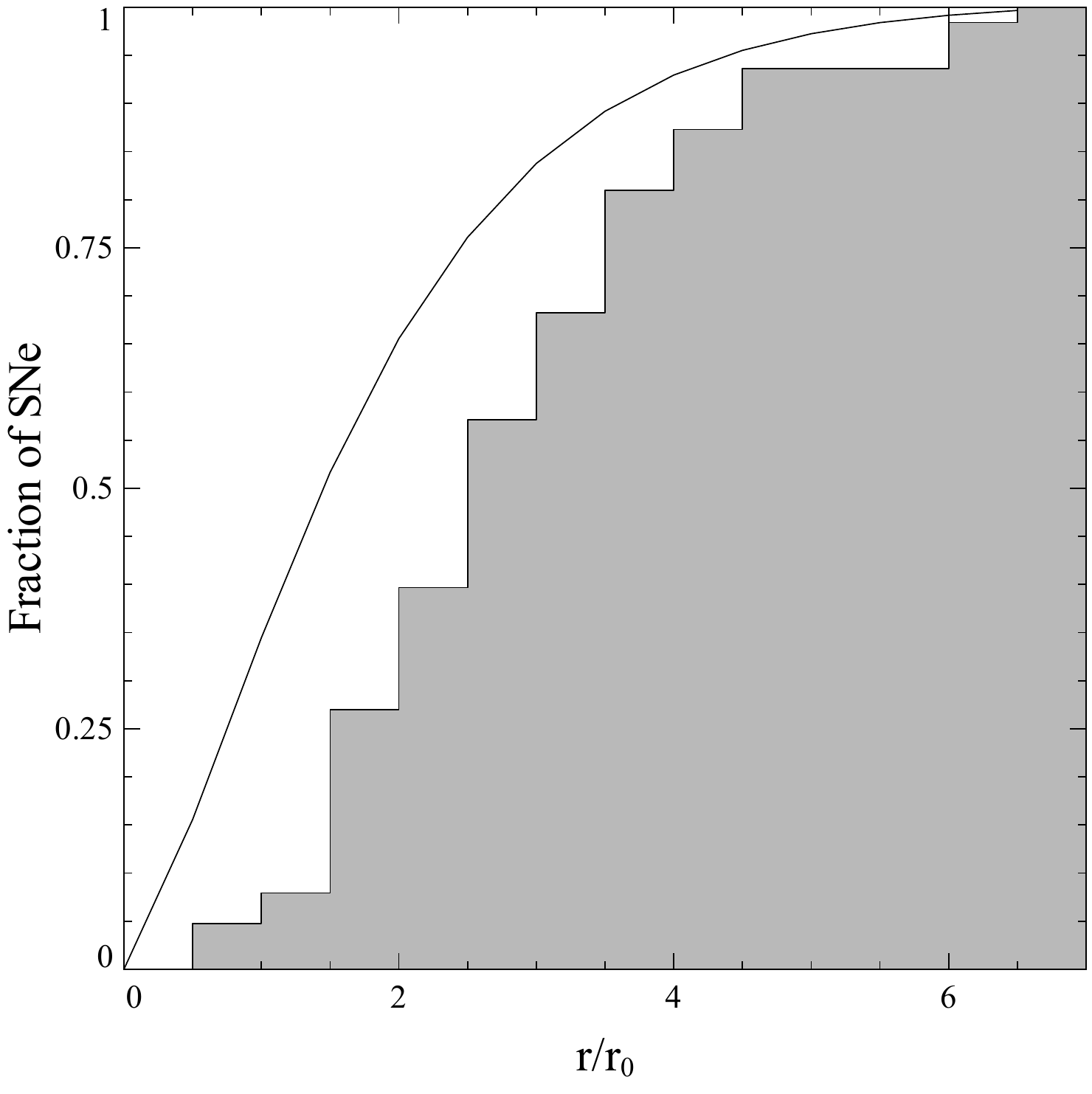}
\caption{A cumulative histogram of the number of SNe Ia vs. radial coordinate showing a deficit at small radii when compared to an exponential density distribution. Note that while the density of stars in a spiral host follows an exponential radial profile, the projected area scales as the square of the radius, resulting in 0 SN counts at the very center.\\}
\label{fig:radial}
\end{figure} 

Note that before the advent of CCD astronomy, it was difficult to spot a supernova against the background light when it occurred near the bright center of its host galaxy (see Shaw 1979). This induces a bias in most SN catalogs, the effect of which is to count fewer supernovae at small radii than would be expected from the stellar distribution in the hosts (e.g.\ Howell \etal 2000).  Figure \ref{fig:radial} shows the radial distribution of SNe Ia in our sample, which indicates a minor deficiency at small radii when our sample is compared to an exponential density distribution. Thankfully, our doughnut method overcomes this bias since it relies only on the local environment of the transient. Constraining our sample, then, to only recently discovered SNe has little to no effect on our results. On the other hand, when global properties are considered, as in a traditional Fruchter \etal (2006) analysis, this bias can have drastic effects on the conclusions about SNe Ia progenitors.

\begin{figure}[!ht]
\centering
\includegraphics[width=7cm,height=7cm]{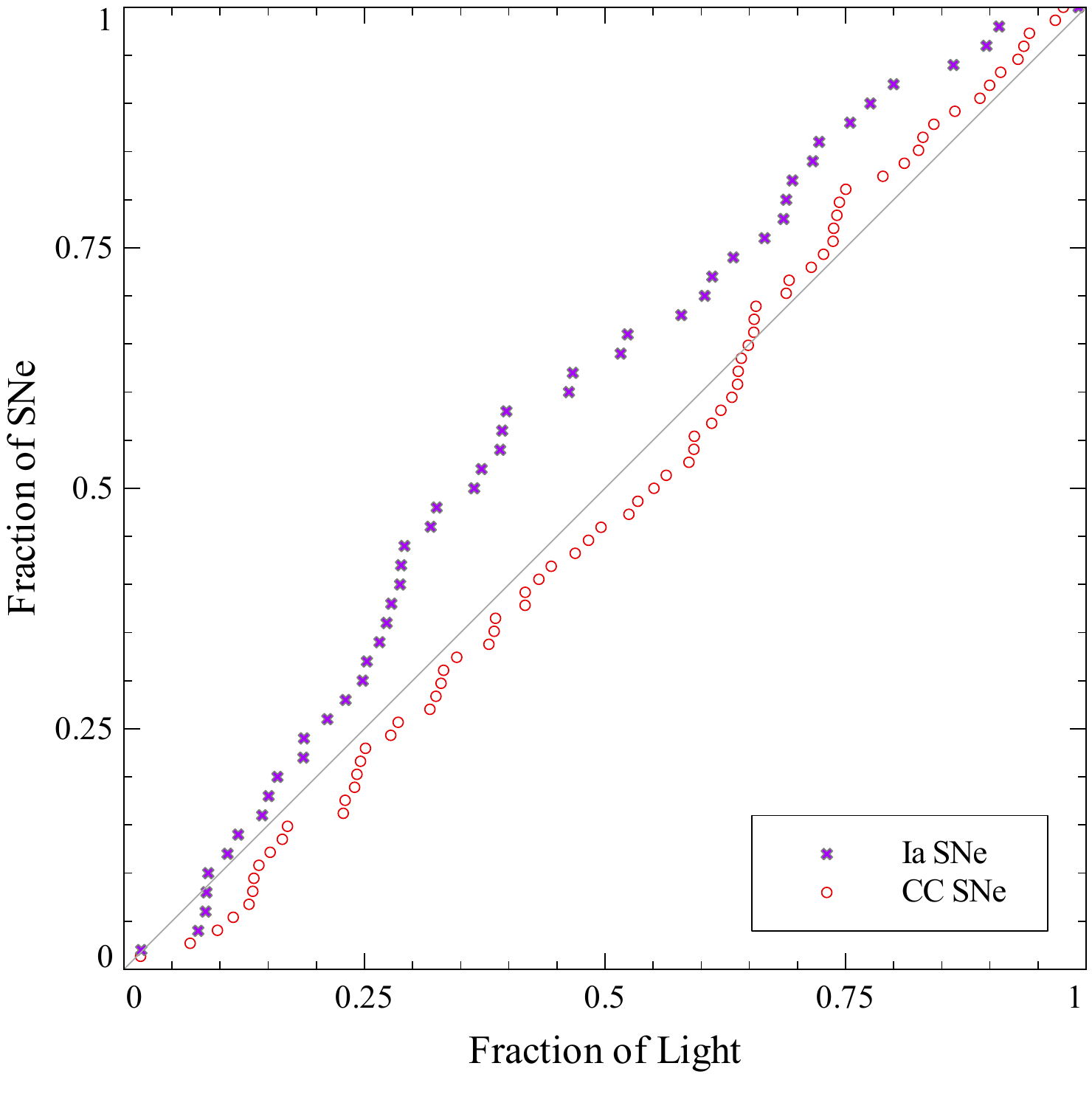}
\caption{A histogram of the number of SNe Ia (crosses) located in pixels brighter than a given fraction of the $g$-band host galaxy light contained within the surrounding ``doughnut,'' which shows that SNe Ia tend to occur in dimmer regions. A histogram of SNcc (circles) is also plotted, which closely follows the $g$-band light distribution.  The two supernova histograms are different at the 99.6\% confidence level.\\}
\label{fig:data}
\end{figure}

Given the typical resolution of SDSS images and the distances to these hosts, we settled on a doughnut width of $0.5r_0$, or $0.25r_0$ on either side of the transient, where $r_0$ is the scale radius.  Host deprojection was carried out using the 25th mag B-band isophote as measured by the Padova-Asiago group, and to obtain the weighted average brightness within the uncertainties in the supernova location, we applied a gaussian convolution with $\sigma = 1.2''$. Figure \ref{fig:data} shows the cumulative supernova distribution vs. the cumulative light distribution. Unlike analyses that use the full galaxy light, our approach is able to distinguish between the SNcc and SNe Ia distributions with 99.6\% confidence, as quantified by a Kolmogorov-Smirnov (KS) test. SNe Ia are clearly biased to fainter pixels, while the SNcc distribution closely follows the $g$-band light distribution.

In order to interpret this separation of the SNe Ia from the SNcc data, we construct an analytical model of a spiral galaxy following that of Raskin \etal (2008) with the necessary modifications to account for the doughnut method. Such a model allows us to age date the SNe Ia by reproducing spatial distributions for a range of SN Ia delay times.

\section{Model Distribution}

In such a model, we define the $g$-band surface brightness at a radius $r$ and angle $\theta$ at a given time, $t$, as the convolution
\be
	\nonumber\Sigma_g(r,\theta,t)=\int_{0}^{t} dt'  L_g(t-t')
	\dot{\Sigma}(t',r,\theta).\label{eq:sfcbrt}
\ee
where $L_g(t)$ is the $g$-band luminosity per unit solar mass of a population stars with a total age of $t$  (Raskin \etal 2008), and $ \dot{\Sigma}(t,r,\theta)$ is the star-formation rate surface density as a function of time and position.  For a galaxy with two spiral arms, this is well-approximated by 
\be
	\dot{\Sigma}(t,r,\theta) = \dot{\Sigma}_0 \exp\left(\frac{-r}{r_0}\right)
	\sum_{n=0}^\infty\delta[\theta-n\pi-\Omega_p(r)t]
	,\label{eq:sigmadot}
\ee
where $\Omega_p$, the pattern speed, is given in eq,\ (\ref{eq:omega}), and $r_0$ is again the scale radius, whence,
\ba
	\Omega_p(r)&=&\frac{v_0}{r_0}\left(1-
	\frac{1}{\sqrt{2}}\right)-\Omega_*(r),\label{eq:omega}\\
\nonumber\Omega_*(r)&=&\cases{v_0/r & $r>\textrm{1kpc}$ \cr
							v_0/\textrm{1kpc} & $r\leq \textrm{1kpc},$\cr}
\ea
with $v_0$ being the circular speed at 1kpc.

In our simulations, we used a pattern speed that rotates $2 \pi$ radians in 300 Myr, although this speed was varied by a factor of two  and ruled out as having a significant effect on the results. Our model is described in  further detail in Raskin \etal (2008) but these details are relatively unimportant as it is the characteristic brightness of the local environment, rather than the angular position of SNe Ia that determine our results.

In order to compute the SN Ia distribution, we replaced $L_g(t)$ with $SNR_{\rm Ia}(t)$ as given by eq.\ (\ref{eq:ratedelay}), where  the prompt component is modeled by a finite-width gaussian distribution of width $\sigma=0.1\tau$ Myr, centered on an adjustable delay time, $\tau$.  Larger values for $\sigma$ were also attempted, but these did not alter the best-fit $\tau$, and instead only reduced the goodness of fit, with the exception of very large $\sigma$ values, which yielded poor fits for all $\tau$. Furthermore, it is crucial that the number of prompt and delayed Ia SNe in our sample be adjustable rather than just adopting the average low-redshift numbers.  This is both becuase our hosts are selected to be star forming, and because of a likely Malmquist bias due to the fact that prompt SN Ia are significantly brighter than delayed SNe Ia. Thus we are left with two independent parameters: $\tau$, the delay time, and $F_{\rm delayed}$  the fraction of delayed SNe Ia in our sample.

\begin{figure}[!ht]
\centering
\includegraphics[width=7cm,height=7cm]{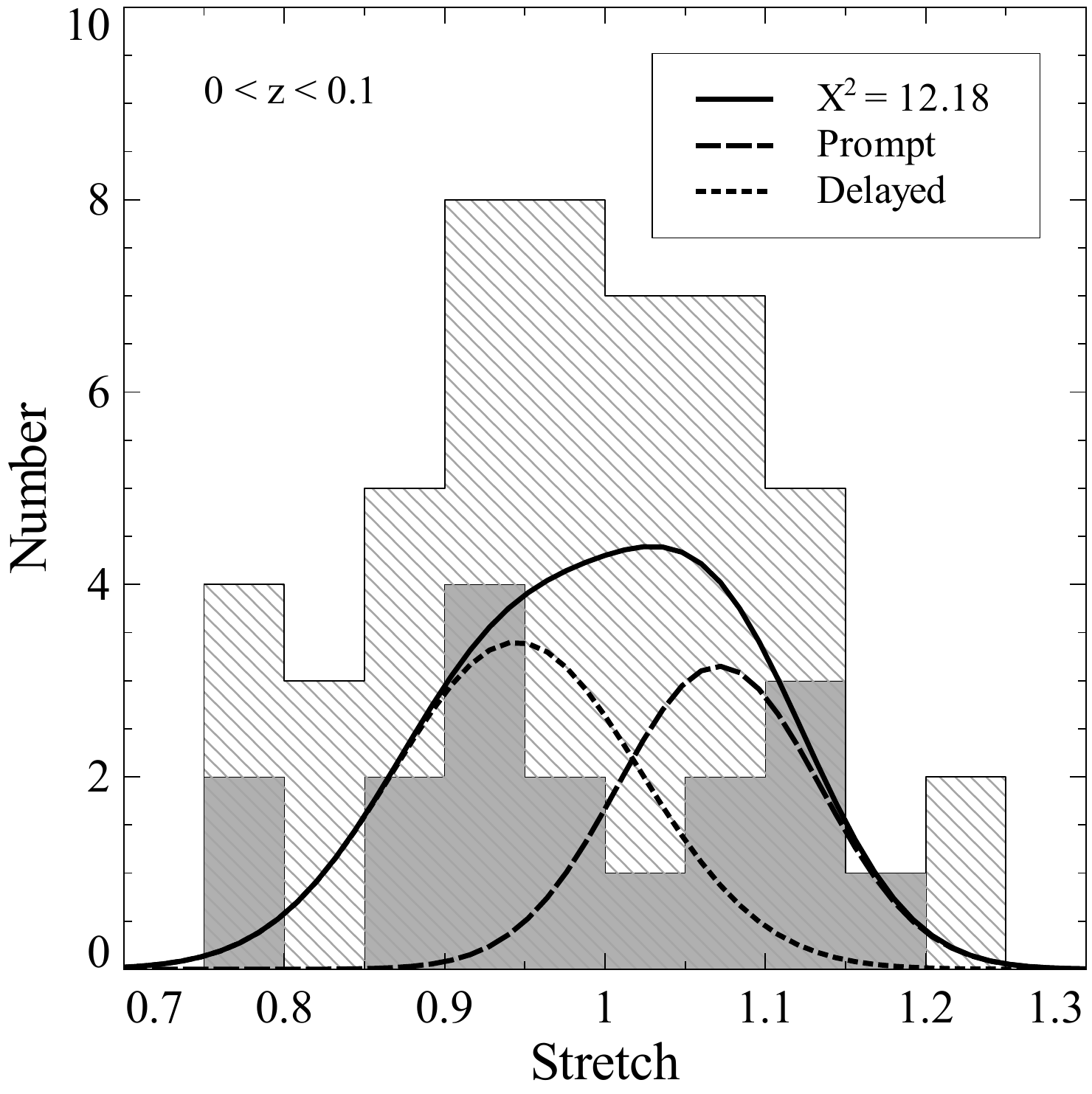}
\caption{A histogram of the distribution of stretches in our sample (dark gray) showing a different distribution as compared to that of Howell \etal (2007) for the same redshift bin (light gray). Using $\bar{s}$ and $\sigma$ values for each component from Howell \etal (2007), a two-gaussian fit for the prompt and delayed components with $\chi^2=12.18$ is overlaid.}
\label{fig:stretches}
\end{figure}

We can estimate this fraction in two ways. First, our SN Ia hosts have a mean color of $B-K = 3.53.$  Within this color bin SNe Ia occur at roughly three times the rate seen in elliptical galaxies (Mannucci \etal 2005), which would suggest that $F_{\rm delayed} \approx 0.3.$ Second, Howell \etal (2007) established a strong correlation between this fraction and the distribution of SN Ia stretches at varying redshifts. They found that long stretches are indicative of prompt SNe Ia, while short stretches indicate delayed SNe Ia. Figure \ref{fig:stretches} shows a histogram of the stretches for a subset of our sample that are available in the literature (dark gray, Conley \etal 2008) and the distribution for the same redshift bin found by Howell \etal (2007, light gray). Again, because our selection criteria favor star-forming galaxies, a resultant two-component fit using the values for mean stretches ($\bar{s}$) and dispersions ($\sigma$) of each component from Howell \etal (2007) demonstrates a mix more heavily favoring the long-stretch, prompt component than would normally be the case in a unbiased sample of low redshift hosts. Integrating these gaussians yields an estimate of $F_{\rm delayed} = 0.56$. For our full analysis then, we consider $F_{\rm delayed}$ to be an adjustable parameter, where our galaxy colors suggest $F_{\rm delayed} \approx 0.3$ and the results of our stretch analysis provide a natural upper limit of $F_{\rm delayed} \le 0.6$. 

For every choice of  $F_{\rm delayed}$ and $\tau$, two images were produced and analyzed using the doughnut method: one representing the host surface brightness and the other representing the SN Ia probability density. The radial light decay across the doughnut was also measured for each observed host. For the most part, $\ln[L(r_{SN}-0.25r_0)]-\ln[L(r_{SN}+0.25r_0)]\approx 0.5$, where $r_{SN}$ is the radial coordinate of the SN, but a few supernovae were found in regions where this relation did not strictly hold, and this variation was used as a small modification to our model in a Monte Carlo fashion. However, the model cannot fully account for the numbers of SNe Ia observed in pixels fainter than the 20th percentile. In the model galaxy, with a doughnut centered on the annulus containing the SN, the dimmest regions will always lie at the outermost edge of the doughnut where SN probability is zero. In real galaxies, dim regions may lie anywhere, and a small number of the SNe Ia will be found in these dim regions.

For the white dwarf formation rate model (WDFR model), $SNR_{\rm Ia}(t)$ was derived from the Chabrier IMF (Chabrier 2003) convolved with the main-sequence turn-off ages from Bruzual \& Charlot (2003) times 10\% to approximate stellar lifetimes, for all stars below 8\msol (see Raskin \etal 2008). Stars whose turn-off ages are older than our simulated galaxy ($\approx$10 Gyr) do not contribute to the simulated SN distribution. Pritchet \etal (2008) suggest a uniform conversion of $\sim$1\% of all white dwarfs to SNe Ia. However, the Fruchter \etal (2006) approach, and by extension our doughnut method, disregards the total count of SNe, instead being concerned only with the relative distribution.

\section{Results}

\begin{figure}[!ht]
\centering
\includegraphics[width=7cm,height=7cm]{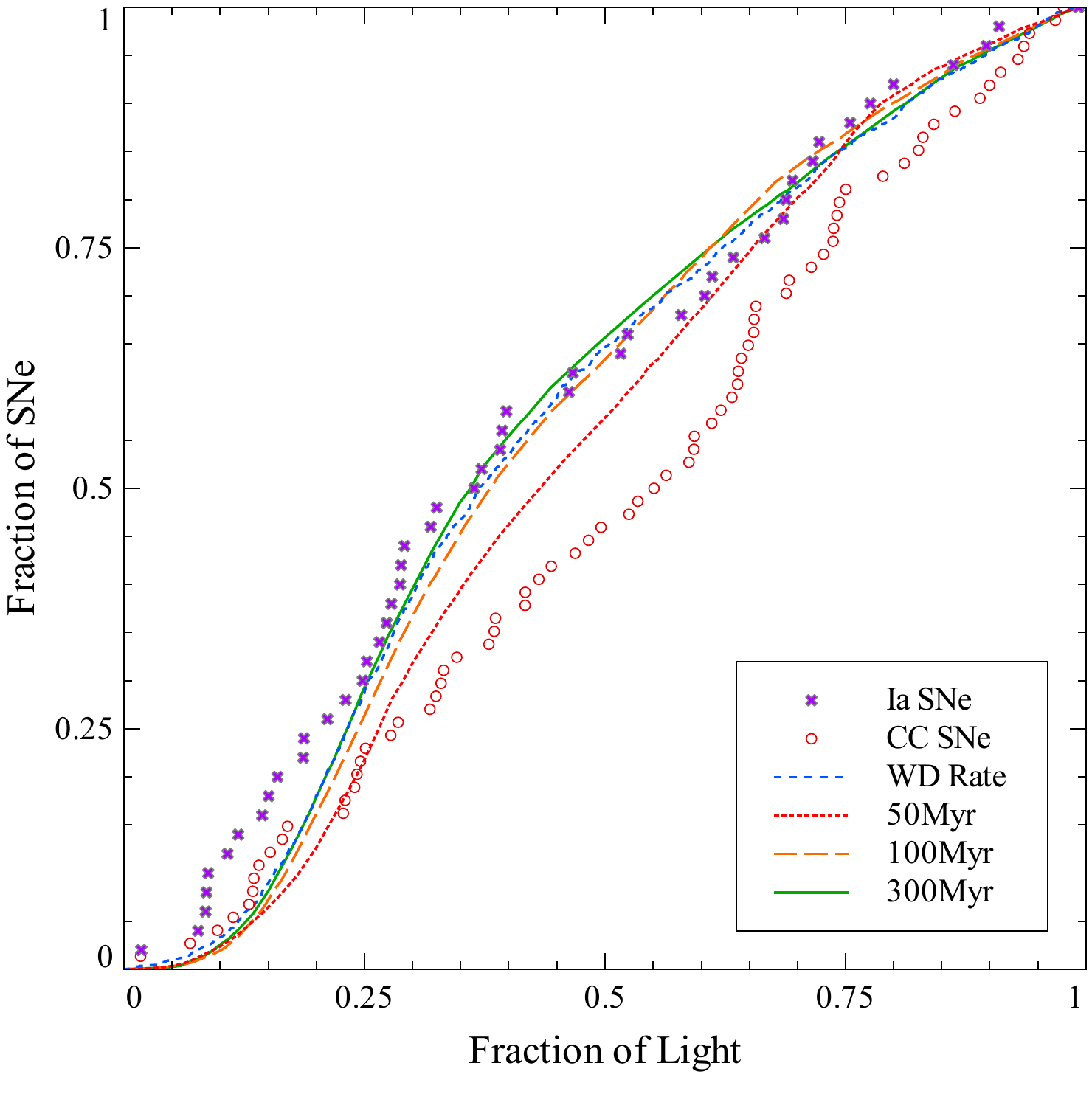}
\caption{Model curves representing the A+B model for several delay times as well as the white dwarf formation rate model are shown compared to the observed distribution of SNe Ia.  Both the $\tau=300$ and WDFR models provide excellent fits to the data.}
\label{fig:fruchter}
\end{figure}

\begin{figure}[!ht]
\centering
\includegraphics[width=7cm,height=7cm]{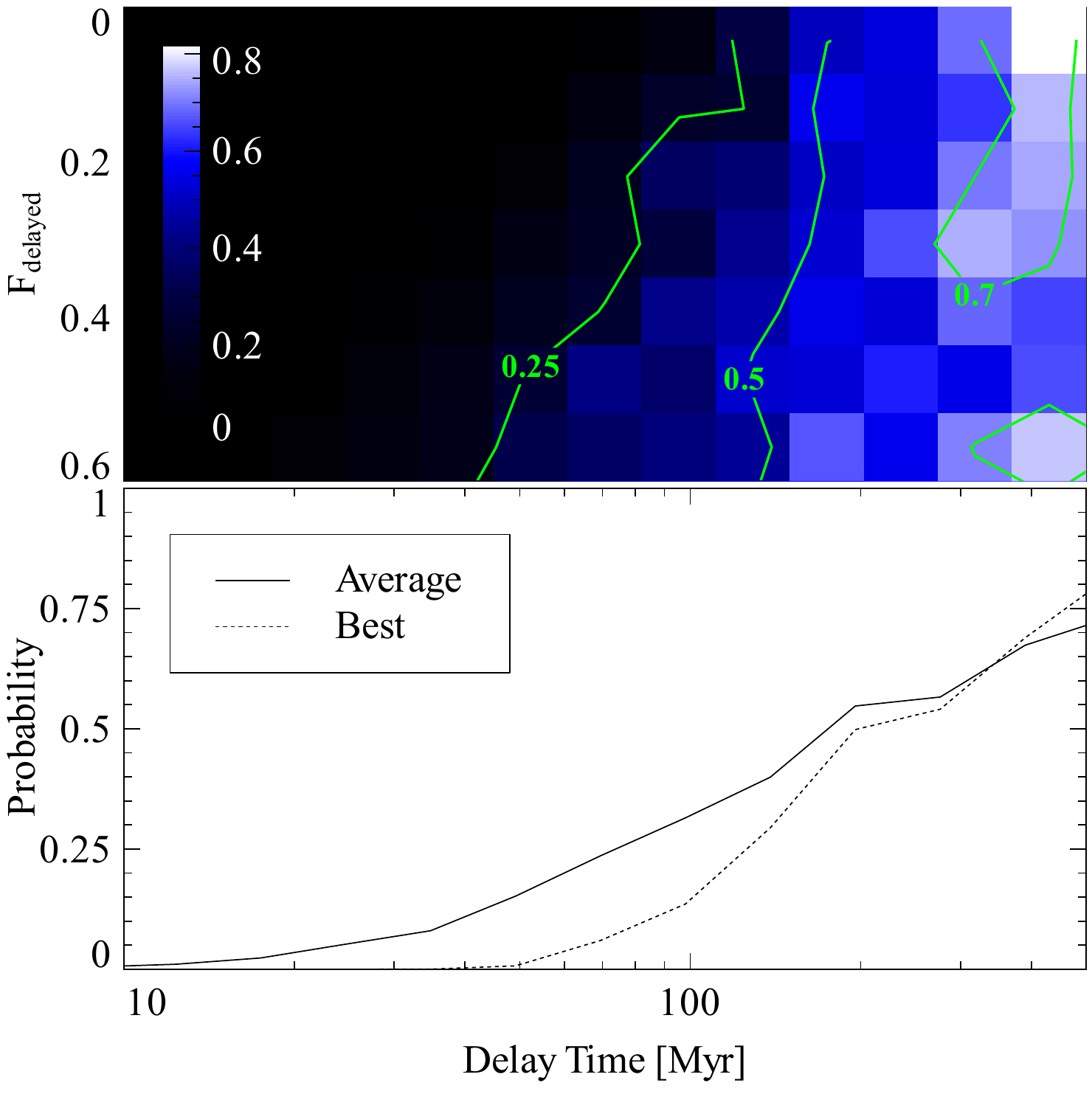}
\caption{\textbf{Top Panel:} A KS probability map for the A+B model with varying parameters $F_{\rm delayed}$ and $\tau$. 
Degeneracies between the two fit parameters are extremely
small. For all values of $F_{\rm delayed},$  $\tau \leq 100$ Myrs is very unlikely, and the
best overall fit value is 500 Myrs. The solid contours mark various levels of KS confidence.
 \textbf{Bottom Panel:}  Average and best fit KS probabilities over a range of $F_{\rm delayed}$ values, each as a function 
 of $\tau,$ again showing that substantial delay times are highly favored. }
\label{fig:ksmap}
\end{figure}

In Figure \ref{fig:fruchter} we plot A+B model distributions with $F_{\rm delayed} = 0.3$ and varying delay times as well as the result of the WDFR model.  It is clear from the A+B model curves that larger values for $\tau$ provide increasingly good fits to the data, while the 50 Myr model is inconsistent with the observations. Because our model galaxy has two free parameters, however, we must account for degeneracy between $\tau$ and $F_{\rm delayed}$. Figure \ref{fig:ksmap} illustrates the two-parameter KS probability map for our sample when compared to our model with varying $F_{\rm delayed}$ values and delay times.

Both the two-parameter map and the average KS probabilities for all tested $F_{\rm delayed}$ values show a clear trend toward longer delay times with the probability reaching its maximum at $\approx 500$ Myr and one value of $F_{\rm delayed}$ yielding a probability of 82\% at this $\tau$.  This is a significant delay time, roughly corresponding to the lifetime of a $2.5\msol$ star at solar metallicity, and is much longer than the minimum theoretically expected time of 40 Myrs. The best fit delay time for all models, even those with high values of $F_{\rm delayed}$, is longer than 200 Myr and probably longer than 300 Myr.   On the other hand, global approaches place an upper limit of $\tau \leq 500$ Myrs (Dahlen \etal 2004; Gal-Yam \& Maoz 2004; Mannucci \etal 2005; Scannapieco \& Bildsten 2005; Sullivan \etal 2006; Dilday \etal 2008). Relating a 200-500 Myr delay to main sequence lifetimes, it is clear that majority of SNe Ia stars have main-sequence masses less than 3\msol, or alternatively, that most SNe Ia arise from  double-white dwarf binaries. It is important to note that while our model provides strong constraints on the {\em characteristic} SNe Ia delay time, it does not rule out that a fraction of SNe Ia occur at shorter delays (Aubourg \etal 2008; Mannucci \etal 2005; Anderson \etal 2007; Anderson \& James 2008; Totani \etal 2008). 

In Figure \ref{fig:fruchter} we also plot the WDFR, whose similarity to the A+B model with $\tau = 300$ Myrs is striking.  One cannot truly be ruled out in favor of the other. A KS test for the WDFR yields a value of 72\%, and in fact, the average delay for this model as calculated by $\int_0^{\infty}{SNR(t)t dt}/\int_0^{\infty}{SNR(t) dt}$ is $\approx 500$ Myr. 

Finally, to place an upper limit on the fraction of prompt SNe Ia with very short delay times we considered  a {\em three-component} model in which $SNR(t) = A M_*(t) + B_1 \dot M_*(t-100 {\rm Myrs}) + B_2 \dot M_*(t-300 {\rm Myrs}).$ By fixing $\tau$, we again consider two free parameters, $F_{\rm delayed}$ and $B_1/B_{Tot}$, the fraction of the prompt component with a delay time of 100 Myrs.  For large values of $B_1/B_{Tot}$, a KS test yields a low statistical likelihood in the model for all values of $F_{\rm delayed}$. We can say with some confidence that the maximum allowed value of $B_1/B_{Tot}$ is 0.3, corresponding to 30\% of the prompt component to coming from a 100 Myr old population, while the parameter space with the highest probability has $B_1/B_{Tot} = 0$.

\section{Conclusion}

Differentiating the spatial distribution of SNe Ia from core-collapse SNe is a difficult problem, both due to the stochastic nature of SN Ia detections and the potential for the delayed component of the Ia rate to obscure the prompt SN Ia spatial distribution.   However, by restricting the Fruchter \etal (2006) analysis to an annulus, our ``doughnut method'' seperates the local host properties from global properties, allowing us to measure the impact of the delay time of the prompt component on the spatial distribution of SNe Ia. By comparing our observations to analytical models, we have established a strong case for a modified A+B model in which the prompt component is  delayed by 300-500 Myrs. Note that this is an average time and that there can be a considerable spread in this value. Thus, a three-component model allows up to 30\% of the prompt component to have shorter delay times $\approx100$ Myrs, as suggested by Auborg \etal (2008) and Mannucci \etal (2006). Alternatively, we also found that a model in which the Ia rate is directly proportional to the white-dwarf formation rate, as considered by Pritchet, Howell \& Sullivan (2008), reproduces the observed spatial distributions of SNe Ia very well.

Regardless of which model proves to be the best fit, it is this characteristic timescale that is most important for calculations of cosmic enrichment, and the results of our analysis using both the A+B model and the WDFR model are evidence of a long characteristic timescale. One caveat is that all our measurements have been made in relatively high-metallicity grand-design sprial galaxies, and several observational and theoretical studies have hinted that SN Ia rates and properties may be substantially different at lower metallicities (Timmes \etal 2003; Gallagher \etal 2008; Cooper \etal 2009). There is still much more to be learned about these ubiquitous but mysterious cosmological probes.

\acknowledgments

This work was supported by the National Science Foundation under grant AST 08-06720.   We thank Don Neill for his help providing supernova stretches for our data, and Filippo  Mannucci, Frank Timmes, Sumner Starrfield, Andy Howell, and our anonymous referee for their many helpful discussions and comments.

\end{document}